\documentclass[twocolumn]{elsart3p}

\usepackage{epsfig}

\begin{document}

\begin{frontmatter}



\title{ Neutron response of the LAMBDA spectrometer and neutron interaction length in BaF$_2$}


\author[label1]{Balaram Dey},
\author[label1]{Debasish Mondal},
\author[label1]{Deepak Pandit},
\author[label1]{S. Mukhopadhyay},
\author[label1]{Surajit Pal},
\author[label1]{K. Banerjee},
\author[label2]{Srijit Bhattacharya},
\author[label3]{A. De},
\author[label1]{S. R. Banerjee\corauthref{cor}}
\corauth[cor]{Corresponding author.}
\ead{srb@vecc.gov.in}

\address[label1]{Variable Energy Cyclotron Centre, 1/AF-Bidhannagar, Kolkata-700064, India}
\address[label2]{Department of Physics, Barasat Government College, Kolkata-700124, India}
\address[label3]{Department of Physics, Raniganj Girls' College, Raniganj - 713358, India}

\begin{abstract}
We report on the neutron response of the LAMBDA spectrometer developed earlier for high-energy $\gamma$-ray measurement. The energy dependent neutron detection efficiency of the spectrometer has been measured using the time-of-flight (TOF) technique and compared with that of an organic liquid scintillator based neutron detector (BC501A). The extracted efficiencies have also been compared with those obtained from Monte Carlo GEANT4 simulation. We have also measured the average interaction length of neutrons in the BaF$_2$ crystal in a separate experiment, in order to determine the TOF energy resolution. Finally, the LAMBDA spectrometer has been tested in an in-beam-experiment by measuring neutron energy spectra in the $^{4}$He + $^{93}$Nb reaction to extract nuclear level density parameters. Nuclear level density parameters obtained by the LAMBDA spectrometer were found to be consistent with those obtained by the BC501A neutron detector, indicating that the spectrometer can be efficiently used as a neutron detector to measure the nuclear level density parameter. 

\end{abstract}

\end{frontmatter}

\section{Introduction}

In the recent past, a Large Area Modular BaF$_2$ Detector Array (LAMBDA) \cite{supm} consisting of 162 BaF$_2$ crystals (each having dimension of 3.5$\times$3.5$\times$35 cm$^3$) has been developed for the measurement of high-energy $\gamma$-rays. These high-energy photons are emitted from the decay of giant dipole resonance (GDR) built on highly excited states of nuclei \cite{harak} as well as from the nucleon-nucleon bremsstrahlung during the early stages of the target-projectile collision \cite{nif}. Until now, the LAMBDA spectrometer has been employed efficiently to reject the neutron contamination from the high-energy $\gamma$-ray spectrum by the time-of-flight (TOF) technique \cite{supm, deep1, supm1, deep2}. However, instead of rejecting the neutrons, they can be utilized to extract the nuclear level density (NLD) parameter which is an inportant ingredient for the statistical model calculation. The most widely used detectors for neutron measurement are liquid hydrocarbon based scintillators such as NE213 \cite{naka} or BC501A \cite{kban}, BC521, BC525 \cite{kban2} etc. because of their good timing and pulse shape discrimination (PSD) properties. However, a BaF$_2$ scintillator can also be efficiently employed for neutron detection because of its excellent timing property (a fast decay component of 0.6 ns) and high density (4.88 g/cc). 

Neutrons having energies E $<$ 10 MeV predominantly interact via (n, $\gamma$) and (n, n$^{\prime}$$\gamma$) reactions with the BaF$_2$ material, whereas for E $>$ 10 MeV, the interaction occurs via different complicated reactions producing charged hadrons \cite{lanza}. The use of a neutron detector usually requires the knowledge of its intrinsic neutron detection efficiency which depends upon many factors, such as, neutron energy, electronic threshold, dimensions of the crystal, interaction mechanism, etc. \cite{kban}. Earlier, many authors have investigated the neutron response of the BaF$_2$ scintillators of various dimensions and over several energy ranges. The efficiency for neutron energies up to 22 MeV was measured by Matulewicz et al. \cite{matu}, while Kubota et al. \cite{kubo} measured the efficiency between 15 MeV and 45 MeV by introducing a PSD cut on the neutron events. The response to fast neutrons in the energy ranges 15-150 MeV and 45-198 MeV was studied by R. A. Kryager et al. \cite{krya} and Gunzert-Marx et al. \cite{marx}, respectively. The investigation was extended to relativistic neutrons with energies up to 1300 MeV by V. Wagner et al. \cite{wagn}. For neutrons in the energy range of 0.5 to 10 MeV, C. Bourgeois et al. \cite{bourg} and Lanzano et al. \cite{lanza} measured the efficiency of  14 cm and 5 cm thick BaF$_2$ crystal, respectively.

In this paper, we report on the neutron response of the existing LAMBDA spectrometer and compare with that of a standard liquid scintillator based neutron detector (BC501A). We show that the neutron energy spectrum measured by the LAMBDA spectrometer in an in-beam experiment can be efficiently used to extract the NLD parameter. We have also measured, for the first time, the average interaction length of neutrons in the BaF$_2$ scintillator to precisely determine the TOF energy resolution which was uncertain in other measurements. Finally, a detailed GEANT4.9.4 simulation has been carried out to understand and explain the neutron response of the LAMBDA spectrometer. 

\begin{figure}
\begin{center}
\includegraphics[height=3.5 cm, width=7.6 cm]{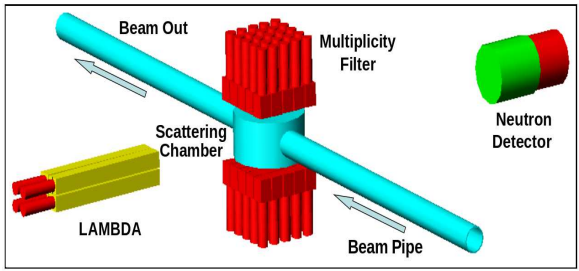}
\caption{\label{setup} \small \sl Schematic view of the experimental set-up.}
\end{center}
\end{figure}

\section{Experimental Details}

\subsection{Efficiency measurement}

The intrinsic neutron detection efficiency of the LAMBDA spectrometer has been measured using a $^{252}$Cf spontaneous fission source (62 $\mu$Ci). $^{252}$Cf decays via $\alpha$ particle emission (96.91\%) and spontaneous fission (3.09\%) with a half-life of 2.65 years and the energy spectrum of the emitted neutron is well documented \cite{knol}. Generally, the fission events are measured using fragment detectors such as surface-barrier detectors \cite{thoen}, PPAC and MWPC \cite{bred}. Since a large number of $\gamma$-rays are emitted from the excited fission fragments, a fast-timing $\gamma$-ray detector (e.g. BaF$_2$) can also be effectively used to select the fission events \cite{deep} as well as to obtain the start trigger for neutron TOF measurement. The number of fission events per second was measured experimentally using the method discussed in Ref \cite{lanza}. A $^{252}$Cf source was placed very close and in front of one of the multiplicity filters (M1) (arranged in 5$\times$5 matrix, discussed later). The other multiplicity filter (M2) was placed at a distance (d) on the other side of the source. The number of coincidence between M1 and M2 was measured for different distances of M2 from the source. For larger distances, a 1/d$^2$ variation of the coincidence rate was observed. For smaller distances, the rate saturated to a value which was consistent with the number of fission events expected from the source activity. The experiment was repeated without the source to reject the background events (e.g. cosmic rays, $\alpha$ impurity in the BaF$_2$ crystal, etc.).
The number of fission per second measured experimentally was used for the efficiency determination. It needs to be mentioned that the measurement contains a very small systematic error ($<$1\%) due to the presence of isotopic impurity in the source.
\begin{figure}
\begin{center}
\includegraphics[height=8.4 cm, width=7.2 cm]{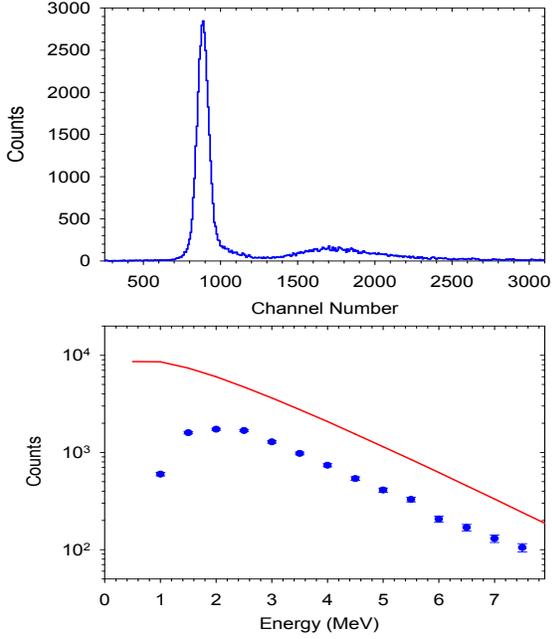}
\caption{\label{fig2} \small \sl [Top panel] TOF spectrum for one of the BaF$_2$ detectors in the array.  [Bottom panel] Neutron energy spectrum of the BaF$_2$ array (filled circle) compared with the expected neutron energy spectrum from $^{252}$Cf (continuous line) \cite{knol}.}
\end{center}
\end{figure}

In the present work, four BaF${_2}$ detectors (a small part of the LAMBDA spectrometer), arranged in 2$\times$2 matrix, were kept at a distance of 80 cm from the $^{252}$Cf source to study the neutron response. The detectors were gain matched and equal thresholds were applied to all of them. A BC501A-based neutron detector (5 inch in diameter and 5 inch in length) \cite{kban} of known efficiency was also employed to measure the neutron energies, in order to compare its efficiency with that of the BaF${_2}$ detectors. The neutron detector was kept on the other side of the source at a distance of 150 cm to equalize the solid angle of the two detector systems. Along with these detectors, a 50 element BaF${_2}$ gamma multiplicity filter \cite{dee} was also used to detect the low energy discrete $\gamma$-rays emitted from the decay of excited fission fragments to establish a correlation between the neutrons and the fission process. The multiplicity filter was split into two blocks of 25 detectors each, in staggered castle type geometry, and placed at a distance of 3 cm above and below the sealed $^{252}$Cf source. A level-1 trigger (A) was generated from the multiplicity filter array when at least one detector from both top and bottom block fired in coincidence above a threshold of 250 keV. Another trigger (B) was generated when the signal in any of the detector elements of the LAMBDA spectrometer or BC501A neutron detector crossed a threshold of 350 keV. A coincidence of these two triggers (A and B) generated the master trigger ensuring the selection of fission events and rejection of background. The schematic view of the experimental set-up is shown in Fig.~\ref{setup}. The TOF technique was employed for neutron energy measurement in both detector systems using the start signal from the multiplicity filter. Along with the time spectrum, the pulse height spectrum of each detector was also measured to apply energy thresholds in offline analysis. A typical TOF spectrum for one of the BaF$_{2}$ detectors (in the array) at a threshold of 350 keV is shown in the top panel of Fig.~\ref{fig2}.

\begin{figure}
\begin{center}
\includegraphics[height=10.0 cm, width=7.2 cm]{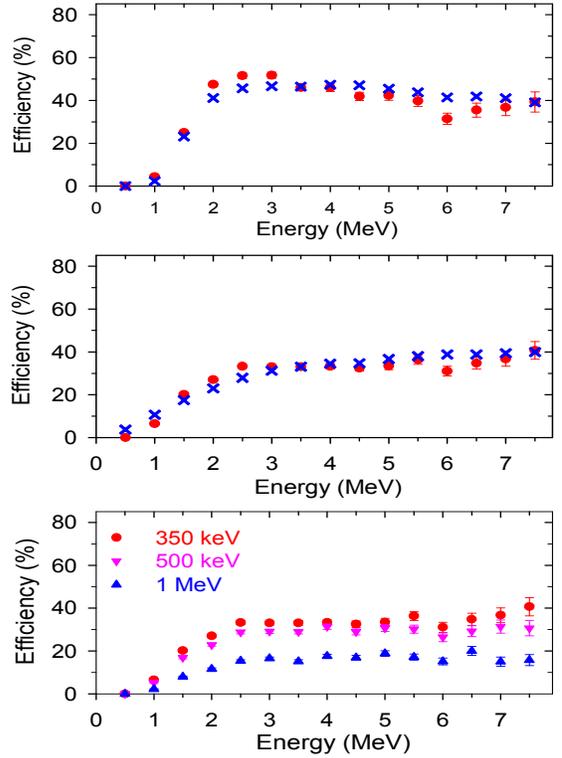}
\caption{\label{eff} \small \sl [Top panel] Energy dependent neutron detection efficiency of BC501A detector. [Middle panel] Energy dependent neutron detection efficiency of BaF$_2$ array. Filled circles are the experimental data points and cross symbols represent GEANT4 simulation. [Botom panel] Neutron detection efficiency at different energy thresholds.}
\end{center}
\end{figure}

The TOF spectrum was converted to energy spectrum using the prompt gamma peak as a time reference. The neutron energy spectrum measured with the BaF$_2$ array (summing all four detectors) is shown in Fig.~\ref{fig2} (filled circles in the bottom panel). The efficiencies of both the detector systems were determined by dividing the neutron yield per fission by the expected neutron energy distribution of $^{252}$Cf \cite{kban,knol} (with temperature T = 1.42 MeV), properly normalized with the detector solid angle and the total number of fission events detected. The efficiencies of individual detectors in the array were also determined. It was found that the intrinsic efficiencies of individual detector elements were identical and very similar to that observed for the array. The energy dependent efficiencies of the BaF${_2}$ array and BC501A detector are shown in Fig.~\ref{eff}. It is interesting to note that, starting at 4 MeV, the neutron efficiency of the BC501A detector decreases monotonically as a function of neutron energy, whereas, the efficiency of the BaF$_{2}$ array increases sharply up to 2-3 MeV and reaches a plateau at efficiency $\sim$34\% which is comparable with that of the neutron detector at these energies. However, the BC501A has an extra advantage in discriminating neutrons from $\gamma$-rays using PSD technique. 

The experimentally measured efficiencies were also compared with the corresponding GEANT4 \cite{gean4} simulation. The GEANT4 simulation for the neutron detector (BC501A) has already been discussed in detail in Ref \cite{kban}. In the case of BaF$_2$ detector, simulation was performed using a series of GEANT4 classes like detector construction and material building, particle and physics process definition, particle tracking, event action, etc. Individual neutrons were randomly generated by a particle generator (G4ParticleGun) and tracked through the detector volume. The energy deposition was recorded step-by-step and finally added for each event. Since neutrons of energies E $<$ 10 MeV interact with the BaF$_2$ material predominantly by (n, $\gamma$) or (n, n$^{\prime}$$\gamma$) \cite{lanza} and deposit their energy in the detector as photons, we considered only the processes of neutron inelastic scattering and capture using G4LENeutronInelastic and G4LCapture models, respectively, in the PhysicsList. We did not include the more complicated nuclear reaction mechanisms and the response of the photo multiplier tube. In the simulation, all the electromagnetic processes were considered for $\gamma$-ray interaction. It was found that the experimentally measured efficiencies of the BaF$_2$ array and BC501A detector were in good agreement with the corresponding GEANT4 simulations. The energy dependent efficiencies of the BaF$_2$ array at various energy thresholds are displayed in Fig.~\ref{eff} (bottom panel). As can be seen from the figure, the overall efficiency decreases with increasing energy threshold, while the nature of the spectrum remains the same. The cross-talk probability of neutrons in the BaF$_2$ array was also estimated in offline analysis and found to be $\sim$12\% at a threshold of 350 keV. It should be mentioned that only statistical errors are shown in the figures. The systematic errors are $<$1\% and arise predominantly due to isotopic impurity present in the source.

\subsection{Time-of-flight energy resolution }

\begin{figure}
\begin{center}
\includegraphics[height=3.8 cm, width=7.0 cm]{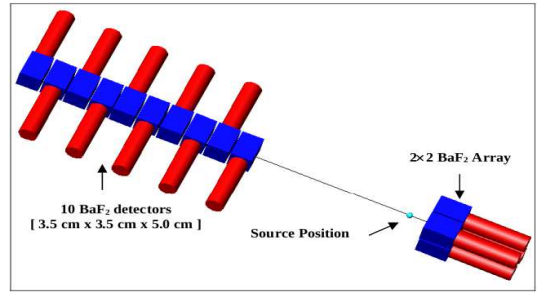}
\caption{\label{setup2} \small \sl Experimental set-up for the measurement of average interaction length of neutrons in BaF$_2$ material.}
\end{center}
\end{figure}

The TOF energy resolution of neutrons is given by the relation, 
\begin{equation}
\left(\frac{\delta E}{E}\right)^2 = \left(2\frac{\delta T}{T}\right)^2 + \left(2\frac{\delta L}{L}\right)^2
\end{equation}
where $\delta E$ is the energy resolution, $\delta T$ is the time resolution of the detector, $L$ represents the mean flight length of the neutron and $\delta L$ is the flight path spread due to the detector size. As the density of the BaF$_2$ material is high, it is expected that, the neutrons will interact mostly in the initial part of the detector volume. Hence, the total size of the detector should not be taken as the uncertainty in length, rather the average interaction length should be estimated and used for energy resolution calculations. The average interaction length of neutrons in the BaF$_2$ detector was measured using the detector elements of gamma multiplicity filter. The length of each detector was 5 cm and its cross-sectional area was same as that of the LAMBDA array elements (3.5 $\times$ 3.5 cm${^2}$). Ten such detectors were arranged linearly (as shown in Fig.~\ref{setup2}) one after another so that the effective length was 35 cm. Next, the detectors were gain matched and equal thresholds were applied to all (300 keV). The TOF spectrum for each detector was measured using a $^{241}$Am-$^{9}$Be source which was kept at a distance of 50 cm from the first of the ten detectors kept in line. The start trigger for TOF measurement was taken from another set of identical BaF$_2$ detectors that were arranged in a 2 $\times$ 2 matrix and kept at a distance of 5 cm on the other side of the source. A schematic view of the experimental set-up is shown in Fig.~\ref{setup2}.

\begin{figure}
\begin{center}
\includegraphics[height=4.5 cm, width=6.5 cm]{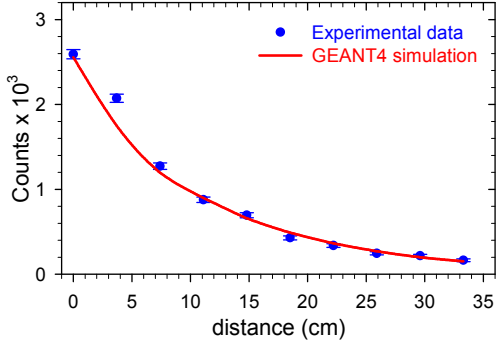}
\caption{\label{neut_interaction} \small \sl Variation of the number of events as a function of distance corresponding to the number of detectors. Filled circles are the experimental data points and continuous line represents GEANT4 simulation}
\end{center}
\end{figure}

In order to estimate the average interaction length, the total number of counts in the neutron TOF spectrum was calculated corresponding to an energy range of 3 - 6 MeV in each of the 10 detectors. Since, a flat overall background was obtained in the TOF spectrum, the background counts were subtracted by selecting the same channels (as for energy bins) from the left of the gamma peak. The total number of neutron events obtained in this energy range (3 - 6 MeV) is shown in Fig.~\ref{neut_interaction} as a function of distance corresponding to the number of detectors. It is very interesting to note that the total number of counts is highest for the first detector and decreases for subsequent detectors, pointing towards the fact that the interaction of neutrons in the BaF$_2$ detector decreases exponentially with increase in distance. A complete GEANT4 simulation was also carried out for this experimental set-up to calculate the average interaction length. As could be seen from Fig.~\ref{neut_interaction}, the experimental data and the simulation results (continuous line in Fig.~\ref{neut_interaction}) match remarkably well with each other. This excellent match between the experimental data and the simulation provided us with the required confidence in GEANT4 simulation. The cross-talk probability was also measured for this set-up and was found to be less than 1\%.

Next, we performed a GEANT4 simulation to estimate the average interaction length for the LAMBDA detector set-up. The interaction points of neutron in the BaF$_2$ material were found to decrease according to the relation exp(-$\mu$x) where $\mu$ = 0.13 cm$^{-1}$. Using this distribution, the average interaction length of  neutrons in the LAMBDA detector was estimated and found to be 7.6 cm when kept at a distance of 80 cm from the source. As a result, the energy resolution at 4 MeV using equation (1) was found to be $\pm$ 0.4 MeV, corresponding to $\delta T$ = 0.96 ns (intrinsic time resolution of the detector).

\section*{3. Measurement of nuclear level density parameter}

The performance of the LAMBDA spectrometer as a neutron detector was tested by measuring the spectrum of the evaporated neutrons in an in-beam experiment. The experiment was performed at the Variable Energy Cyclotron Centre, Kolkata using a 35 MeV alpha beam from the K-130 cyclotron. A self-supporting foil of $^{93}$Nb (99.9\% pure) with a thickness of $\sim$1 mg/cm$^{2}$ was used as the target. The compound nucleus $^{97}$Tc* was populated at the initial excitation energy of 36 MeV. The experimental set-up was similar to that used for the efficiency measurement and is shown in Fig.~\ref{setup}. To keep the background of the detectors at a minimum level, the beam dump was kept 3 m away from the target and was well shielded with the layers of lead and borated paraffin. Data from the BaF$_2$ and BC501A detectors were recorded in an event-by-event mode in coincidence with the $\gamma$-multiplicities in order to measure the neutron energy spectrum and to extract the angular momentum of the compound nucleus. The TOF technique was employed for neutron energy measurement in both the detectors using the start trigger from the  multiplicity filter. Along with the time spectrum, the pulse height spectrum of each detector was also measured to apply the energy thresholds in offline analysis. The cross-talk probability of neutrons in the BaF$_2$ array was estimated in the above experiment and found to be same as that obtained in the efficiency measurement with $^{252}$Cf.

\begin{figure}
\begin{center}
\includegraphics[height=8 cm, width=7.0 cm]{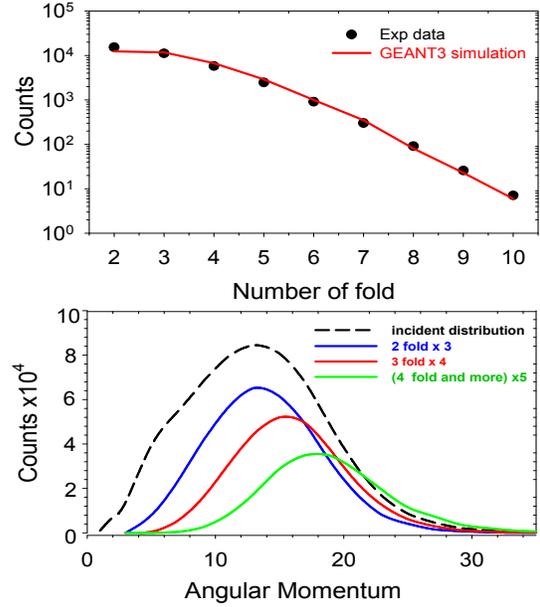}
\caption{\label{fold} \small \sl [Top panel] Measured fold distribution along with the GEANT3 simulation fit. [Bottom panel] Angular momentum distribution for different folds for the $^{4}$He + $^{93}$Nb system}
\end{center}
\end{figure}

\begin{figure}
\begin{center}
\includegraphics[height=10.5 cm, width=6.8 cm]{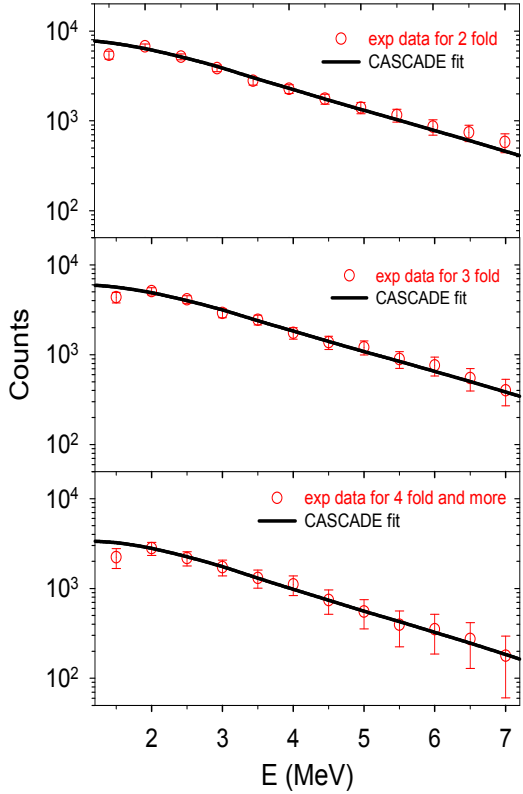}
\caption{\label{bc501acas} \small \sl Experimental neutron energy spectra (open circles) of BC501A detector along with the CASCADE fit (continuous line) for different folds.}
\end{center}
\end{figure}

\begin{figure}
\begin{center}
\includegraphics[height=10.5 cm, width=6.8 cm]{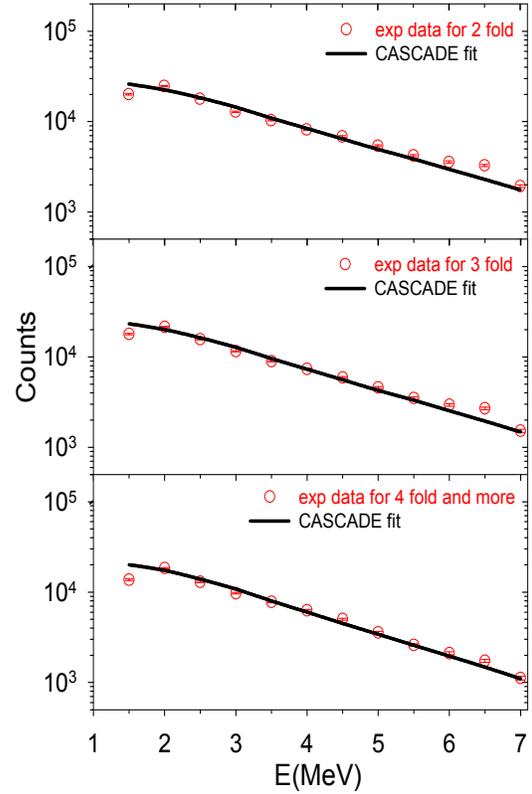}
\caption{\label{bafcas} \small \sl  Experimental neutron energy spectra (open circles) of BaF$_2$ detector along with the CASCADE fit (continuous line) for different folds.}
\end{center}
\end{figure}

The experimental fold distribution measured using the 50-element $\gamma$-multiplicity filter is shown in the top panel of Fig.~\ref{fold}. The fold distribution was converted to the angular momentum distribution applying the approach discussed in Ref \cite{dee}. The angular momentum distributions corresponding to different folds are shown in the bottom panel of Fig.~\ref{fold} while the average values are given in Table 1. The neutron energy spectra were extracted from the TOF spectra using the prompt gamma peak as a time reference. The neutron energy spectra measured using the BC501A and BaF$_2$ array are shown in Fig.~\ref{bc501acas} and Fig.~\ref{bafcas}, respectively (open circles).

The asymptotic level density parameter $\widetilde{a}$(A) = ${A}/{k}$ is an important input for CASCADE calculation \cite{cas}, where k is kept free and generally adjusted to get the best fit with experimental data. We extracted the values of k from the neutron energy spectra measured using the BC501A and BaF$_2$ array for different folds of the multiplicity filter. The simulated angular momentum distribution corresponding to each fold was used as input in the modified version of the statistical model code CASCADE. The value of k was extracted from the experimental data by chi-square minimization technique (3-7 MeV). The experimental neutron energy spectra along with the CASCADE predictions for the BC501A and BaF$_2$ detector systems are shown in Fig.~\ref{bc501acas} and Fig.~\ref{bafcas}, respectively. The best fit values of the level density parameter (k) for different folds are displayed in Table 1. It is interesting to note that level density parameters extracted from the BaF$_2$ and BC501A detectors are in good agreement. It is also observed that the values of k decrease with the increase in angular momentum similar to the results obtained earlier for charged particle and neutron measurements \cite{proy} from the same system. Thus, the excellent match between the level density parameter obtained from the BC501A and BaF$_2$ array clearly suggests that the LAMBDA spectrometer can be effectively used to measure the neutron evaporation energy spectrum along with the high energy $\gamma$-ray spectrum.


\begin{table}
\caption[]{\label{tab:nucl} \small \sl The values of k corresponding to different folds of BaF$_2$ and BC501A detectors}
\begin{center}		
\begin{tabular}{|c|c|c|c|}
\hline
Fold     & Angular momentum &  k (MeV)   & k (MeV) \\
      &	$\left\langle J \right\rangle$$\hbar$  & (BaF$_2$ array)  & (BC501A) \\
\hline
2 &     14    &    10.8 $\pm$ 0.4  &    10.4 $\pm$ 0.2   \\
\hline
3 &     16    &    10.5 $\pm$ 0.4  &  10.3 $\pm$ 0.4     \\
\hline
 4 $\&$ more  &  19  &  9.6 $\pm$ 0.3  &  9.7 $\pm$ 0.3   \\
\hline
\end{tabular}
\end{center}		
\end{table}

\section*{4. Summary and Conclusion }

In summary, we have measured the energy dependent intrinsic efficiency of the LAMBDA spectrometer for neutron detection relative to a liquid organic scintillator based neutron detector (BC501A) of known efficiency and compared with a detailed GEANT4 simulation. In this work, we have also determined the average interaction length of neutrons in the LAMBDA spectrometer to estimate the time-of-flight energy resolution. Furthermore, we tested the spectrometer in an in-beam experiment ($^{4}$He + $^{93}$Nb) to measure the nuclear level density parameter and compared with that measured using the standard BC501A neutron detector. Interestingly, it was observed that the values of k obtained from both the detector systems are very similar, indicating that the BaF$_2$ detector can be efficiently used for the measurement of evaporated neutrons from an excited compound nucleus in an in-beam experiment along with its intended use in the measurement of high energy $\gamma$-rays.



\end{document}